\documentclass[12pt]{article}

\usepackage{amsmath,geometry,amsthm,amssymb,natbib,verbatim,upref,paralist,indentfirst,verbatim,graphicx,pdfpages,caption,hyperref,enumitem}

\theoremstyle{definition}

\usepackage[T1]{fontenc}
\usepackage[utf8]{inputenc}
\usepackage{authblk}
\usepackage{tgheros}
\begin{document}

\title{Blockchains, MEV and the knapsack problem: a primer\thanks{This research is funded by The Ethereum Foundation Academic Grant Program no. FY22-0682. The authors thank Barnabé Monnot for useful discussions and feedback on the content and implementation of this grant project.}}

\author[1]{Vijay Mohan\thanks{Corresponding author; email: vijay.mohan@outlook.com.}}
\author[2]{Peyman Khezr}

\affil[1]{\textit{\footnotesize Lattice Analytics Pty Ltd, Melbourne, Australia.}}
\affil[2]{\textit{\footnotesize School of Economics, Finance and Marketing, RMIT, Melbourne, Australia.}}

\date{}

\maketitle

\begin{abstract}

In this paper, we take a close look at a problem labeled \textit{maximal extractable value} (MEV), which arises in a blockchain due to the ability of a block producer to manipulate the order of transactions within a block. Indeed, blockchains such as Ethereum have spent considerable resources addressing this issue and have redesigned the block production process to account for MEV. This paper provides an overview of the MEV problem and tracks how Ethereum has adapted to its presence. A vital aspect of the block building exercise is that it is a variant of the knapsack problem. Consequently, this paper highlights the role of designing auctions to fill a knapsack—or \textit{knapsack auctions}—in alleviating the MEV problem. Overall, this paper presents a survey of the main issues and an accessible primer for researchers and students wishing to explore the economics of block building and MEV further.

\end{abstract}

\vspace{1ex}





\section{Introduction} \label{Sec1}

In a blockchain, transactions submitted by users are grouped together to form a block. This block is then verified by the network and attached to a previous block, thereby creating a time-ordered chain of linked blocks. We refer to the entity responsible for creating blocks as a \textit{block producer}.\footnote{The term ‘block producer’ is often used in the context of proof-of-stake systems. We use it more generally here to denote any entity that produces a block because, no matter what the underlying consensus system is, blocks have to be created by some agent on a blockchain.} Depending on the consensus mechanism of the blockchain, the block producer may be a \textit{miner} in the case of proof-of-work (PoW), or a \textit{validator} in proof-of-stake (PoS).\footnote{Proof-of-work is the original consensus mechanism proposed by \cite{nakamoto2008bitcoin}, wherein miners create blocks by solving a complex mathematical puzzle. In a proof-of-stake mechanism, agents stake tokens to receive the right to be selected randomly to produce a block. For an introduction to consensus mechanisms, see \url{https://ethereum.org/en/developers/docs/consensus-mechanisms/} (accessed 6th March 2024).} In exchange for creating the block, producers are entitled to compensation which, in general, can take the form of \textit{block rewards} and \textit{fees}. The former refers to the creation of new tokens in the blockchain to pay for block creation. Bitcoin miners, for example, earn block rewards of 6.25 BTC per block at the time of writing, and PoW miners on Ethereum historically (prioir to 2022) earned 2 ETH.\footnote{Bitcoin block reward changes over time and becomes half after a certain period of time till the maximum supply of 21 million BTC is reached. As expected, block rewards are paid in a blockchain’s native token. The current reward scheme on Ethereum after the move to PoS in 2022 is described in Section 3.} Fees are added by users to ensure that their transactions receive priority for inclusion in a newly created block. In this paper, we summarize how the decision-making process of a block producer and those who want to place their transactions into a block relates to a well-known problem called the \textit{knapsack problem}.\footnote{See \cite{bartholdi2008knapsack} for the basic knapsack problem and \cite{kellerer2004multidimensional} for an advanced treatment.} In doing so, we present an overview of some important research issues related to the block production process that currently confronts the blockchain community.  

In what follows, we structure our discussion primarily around the Ethereum ecosystem to develop our ideas further. Ethereum currently operates on a PoS consensus mechanism. However, we also relate parts of our discussion to Bitcoin and earlier iterations of Ethereum, both of which incorporated a PoW consensus mechanism. Ethereum not only allows transactions to be conducted by users - like Bitcoin does - by utilizing externally-owned accounts (EOAs) that are controlled by private keys, but also deploys contract accounts or \textit{smart contracts} that are controlled by code. As such, Ethereum serves as a virtual computer where users can implement code through smart contracts.\footnote{See \url{https://ethereum.org/en/developers/docs/accounts/} (accessed 6th March 2024).} Since the computations involved with implementing code utilizes resources, Ethereum measures the amount of computational effort using a unit referred to as \textit{gas}. 

Transactions are conveyed to block producers through a \textit{mempool}, which essentially is a collection of transactions waiting to be included in a block.\footnote{Technically, there is no single mempool as each node in the network receives transactions at different times, so that different nodes may have a different set of transactions waiting to be included in a block. We abstract from this feature here and assume that there exists a single mempool common to all nodes on a network, which simplifies the exposition without impacting the main thrust of the ideas presented here.} The mempool is public, and all agents on the blockchain can see the transactions contained in it.\footnote{Given that the mempool is public, one could refer to it as a demand schedule, where potential consumers state their quantity demanded and their maximum willingness to pay.} In general, users also have the option of submitting transactions privately to block producers, but in earlier iterations of Ethereum this required directly interacting with block producers, which is a difficult proposition for the average user.  As we shall elaborate upon presently, the ability to submit private transactions has evolved in Ethereum and is part of a design choice aimed at tackling a specific problem. Figure \ref{fig1} below depicts the overall system in its simplest form. In the first step, users submit transactions through the mempool, which is the input into the system. The block production process can vary depending on the nature of the consensus and the institutional structures created by a specific blockchain community. But no matter how it is constructed, the ultimate output of the system are blocks of transactions, which are concatenated to create a blockchain. The flow of transactions is shown in blue in Figure \ref{fig1}, while the flow of value and funds is shown in red. 

\begin{figure}[h!]
\centering
  \includegraphics[scale=0.8]{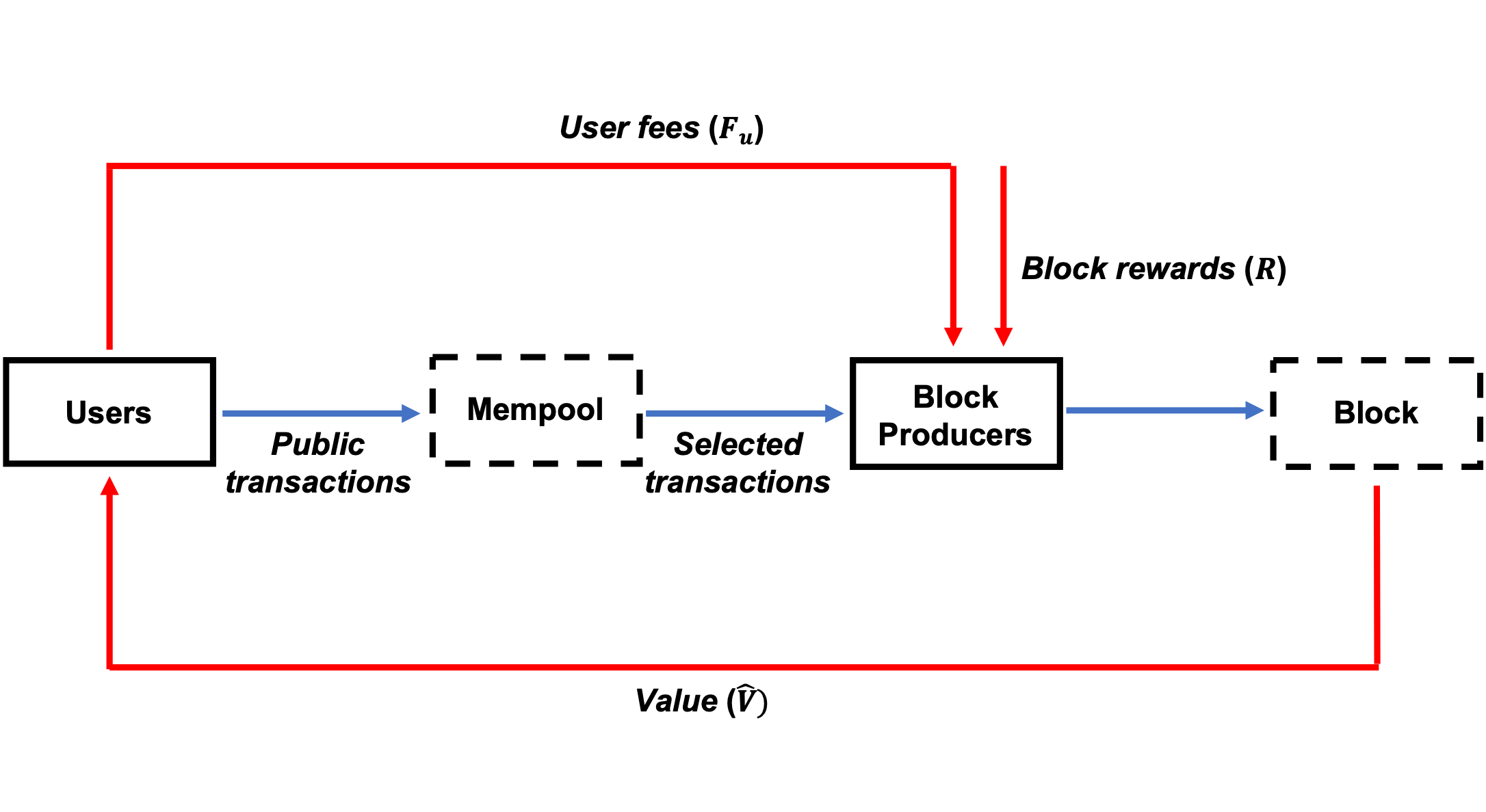}  
\caption{The flow of funds, transactions and value on a blockchain}
\label{fig1}
 \end{figure}

A key element here is \textit{value}, which summarizes the benefits that users receive from a block being created using their transactions. Value could represent, for example, the consumer surplus that a user derives from the purchase of a coffee using bitcoin, or the profits that can be made from an arbitrage opportunity observed on Ethereum. In Figure \ref{fig1}, the total value is $\hat{V}$, and all of it flows back to the users who submitted the transactions. As we shall see in Section \ref{Sec3}, one of the major problems confronting Ethereum and other similar platforms is the possibility that value is transferred away from users who originally submit transactions due to a phenomenon that has been labelled \textit{maximal extractable value} (MEV). Block producers are rewarded with fees paid by users ($F_u$) and block rewards ($R$) for creating and validating blocks. Block rewards involve the creation of new tokens, much like `helicopter money', and are shown in Figure \ref{fig1} as an exogenous injection into the system from outside. Based in the flows of value and funds in Figure \ref{fig1}, we can track the payoffs to users ($\pi_u$) and block producers ($\pi_b$). 

\begin{equation}\label{eq1}
\begin{aligned}
& \pi_{u} = \hat{V} - F_{u}
&&\\
& \pi_{b} = F_{u} + R 
&&
\end{aligned}
\end{equation}   

\noindent The \textit{total surplus} from the creation of a block is the sum of the two payoffs:

\begin{equation}\label{eq2}
\begin{aligned}
&\Pi = \pi_{u} +  \pi_{b} = \hat{V} + R\\
\end{aligned}
\end{equation}

A block has a finite size and can therefore fit only a finite set of transactions from the set of available transactions. Consequently, a major task of the block creation process is the selection of the transactions that go into a block. If all transactions were the same size and carried the same fees then, all else being the same, the block producer would not care which transactions from the mempool are selected for building a block. There are, however, three reasons that make this task non-trivial in reality. First, transactions have different sizes. For example, Bitcoin transactions can vary in size based on how many inputs and outputs they have;\footnote{See \url{https://en.bitcoin.it/wiki/Scalability_FAQ\#Should_miners_be_allowed_to_decide_the_block_size.3F} (accessed 6th March 2024).} similarly, on Ethereum, which employs smart contracts, transactions vary in terms of the amount of gas involved for computation. Secondly, as mentioned earlier, transactions vary in terms of the fees offered by users to block producers for prioritizing the transactions selected into a block. Third, the structure of the decision-making process that the block producer must undertake is a knapsack problem, which is known to be an NP-hard problem. While we focus on blockchains and MEV in this paper, many of the ideas presented here translate to other contexts as well, such as allocating advertisements in a fixed interval of time, allocating cargo space to different objects, selecting groups to fit in a stadium during a sporting event, and so on. 

The remainder of the paper is organized as follows. Section \ref{Sec2} takes a look at how a block is packed with transactions. The focus here is on describing the knapsack problem, and how it translates to a market design problem. Section \ref{Sec3} presents a primer on MEV, using Ethereum to provide context to the analysis. Ethereum has undergone numerous changes in the last few years, some of which have been undertaken to tackle MEV. Consequently, this section reviews relevant changes in the Ethereum network as well. Section \ref{Sec4} provides an overview of the role played by auctions in the execution of the knapsack problem in the context of blockchains, which translates to the design of knapsack auctions. Section \ref{Sec5} offers some concluding comments.

\section{The knapsack problem}\label{Sec2}

\subsection{The knapsack problem as a decision-theoretic problem} \label{Sec2.1}

At a very broad level, the knapsack problem deals with a sequence of binary decisions regarding the usage of a scarce resource.  Consider, for example, a hiker or mountaineer who has a knapsack that needs to be filled with objects before an expedition, or a salesman who has to pack a suitcase before travelling to different cities, or a Bitcoin miner who has to pack a block with transactions. In all these cases, the fundamental problem is that there is a container with a fixed capacity (knapsack or suitcase or block) and the objects that need to be fitted into the container exceeds the capacity.\footnote{If the knapsack can fit all the desirable objects, then there exists no meaningful problem: all the objects will be packed.} So, the mountaineer may be limited by the space available in the knapsack or the weight that can be carried in the expedition (or both) and the block-producer by the amount of data that can be included in a block (Bitcoin) or the gas usage in a block (Ethereum). Given that all the objects cannot be packed, the decision-maker has a binary choice for each object: whether to pack it or not. Each object varies along two dimensions: its size and its value. The former refers to how much of the scarce resource (say space in the knapsack) the object utilizes when it is packed, and essentially translates to the cost of packing the object; the latter relates to the benefit the object provides to the decision-maker when it is packed. Couched in this way, solving the knapsack problem is nothing but a cost-benefit analysis when utilizing a scarce resource.

Economists are, of course, familiar with this problem in various forms. One could covert, for example, the decision-making process of a consumer into a knapsack problem. The consumer has a budget with a fixed capacity, and the consumer must select the goods and services that need to ‘packed’ into this budget when each object has an associated size (price) that utilizes a scarce resource (money) in order to produces value (utility) for the consumer. But maximizing utility subject to a budget constraint is something that economists routinely do, so one may well ask why the knapsack problem deserves any attention in this context. The answer, as expected, has to do with the assumptions that economists make to neatly side-step the complexity of this problem. Specifically, consumer theory usually assumes that the utility function is continuous, and that objects are perfectly divisible. These assumptions allow the use of calculus to find a solution to the optimization problem, which simply involves maximizing the bang-for-buck from each dollar spent, and yields an equilibrium condition where the marginal utility per dollar spent is equated across goods. 

Carrying over the economic intuition from consumer theory back to the knapsack problem, one can begin with the question of what happens if the objects that need to be fit into the knapsack were perfectly divisible and one followed the bang-for-buck approach by packing objects in decreasing order of value per size till the last object is reached, which can then be divided into whatever portion is required to fill the final bit of vacant space in the knapsack. This method, labelled the \textit{Greedy algorithm} \cite{dantzig1957discrete}, does in fact fill the knapsack in a manner that maximizes the total value of the packed objects. So, when objects are divisible, the solution to the knapsack problem is as simple as the textbook consumer optimization problem.

In some cases, divisibility is reasonable, for example a kilogram of rice can be divided to the extent of a single grain. In many other instances, such as white goods, furniture and printers, objects are non-divisible and the number of objects packed must assume integer values. While economists have tended to ignore the complexity that this causes, computer scientists have not. Indeed, the knapsack problem with non-divisibility (often called the 0-1 knapsack problem) is considered an NP-hard problem, which implies that there is no known algorithm that can solve the problem in polynomial time \citep{kellerer2004multidimensional}. In all the examples given above ---the mountaineer, the travelling salesman, the block producer, TV advertisements--- the objects typically cannot be subdivided in a meaningful way. In a blockchain, a transaction is either fitted in its entirety in a block or it is not, making it a 0-1 knapsack problem. The block production exercise is, therefore, NP-hard; the complexity can be visualized by observing that the mempool can have thousands of transactions waiting to be included in a block, from which block producers have to select a subset in a manner that satisfies their objectives (usually maximizing the fees obtained from constructing a block). As a result, block producers typically use the Greedy algorithm that only provides some approximation of the optimal outcome as objects cannot be divided to occupy all the space in the knapsack.\footnote{The Greedy algorithm can fail to produce an optimal outcome. For example, consider an extreme case where there is a knapsack of capacity $K$ and there are only two objects; the first has size and value both equal to $1$ and value $1$, while the second has size $K$ and value $K-\delta$, where $\delta > 0$. The value to size ratio of the first object is $1$, and of the second object is $(K-\delta)/K  <1$. If one follows the Greedy algorithm and packs according to the value to size ratios, the first object is packed, which leaves no space to fit the second object, which is then excluded. This is clearly sub-optimal when $K - 1> \delta$.}

In the knapsack problem described above, we have assumed there is a single decision maker who knows both the size and value of the objects and consequently acts in an environment of complete information. This is realistic, for example, when the decision-maker owns both the knapsack and the objects that need to be fit into it. Consequently, it is relevant for economic situations such as a company that owns both a vehicle and the objects transported in it, or indeed a consumer seeking to allocate a fixed budget across goods. Below we summarize the elements that comprise the basic knapsack problem with complete information. 

\vspace{0.25cm}

\noindent \textit{\textbf{Model I: The general knapsack problem with a single agent and complete information}} 
\begin{enumerate}[nosep, leftmargin=*]
\item A set of $N=\{1, 2, 3,..., n\}$ \textit{objects}.
\item There exists a single agent, and is responsible for making all decisions related to packing the knapsack.
\item A knapsack of \textit{capacity}, $K>0$, into which the objects must be packed.
\item Each object $i\in N$ has a \textit{size} of $k_{i}>0$ associated with it.
\item Each object $i\in N$ has a \textit{value} associated with it, which accrues to the agent if the object is fitted into the knapsack; if not, the object has zero value. 
\item The agent knows the size ($k_i$) and value ($v_i$) of each object, making this a problem of complete information.
\item A decision process, $x_i$, that determines which objects are are placed in the knapsack. There are two variations:
\begin{enumerate}[nosep]
    \item The fractional problem where $x_{i} \in [0,1], \forall i \in N$; in this variation an object is divisible and any fraction can be packed in the knapsack.
    \item The binary problem where $x_{i} \in \{0,1\}, \forall i \in N$; in this variation an object is not divisible and is either placed in the knapsack ($x_{i}=1$) or not ($x_{i}=0$). 
\end{enumerate}

\item The objective of the decision-maker is to maximize the sum of values or the \textit{surplus} obtainable from filling the knapsack:

\begin{equation}\label{eq3}
\begin{aligned}
& \underset{x_{1},x_{2},...x_n}{\max}
&& \sum_{i=1}^{n} x_{i}v_{i} \\
& \textit{subject to} 
&& \sum_{i=1}^{n} x_{i}k_{i}\leq K \\
\end{aligned}
\end{equation}   

\end{enumerate}

\noindent If 7(a) holds, the Greedy algorithm attains the maximum surplus $\sum_{i=1}^{n} x_{i}v_{i} = V^*$. But this is unrealistic in our context of blockchains, as a transaction is either included in a block ($x_i=1$) or not ($x_i=0$). Consequently, we assume that 7(b) holds in Model I. In that case the program described by Equation \ref{eq3} is NP-hard and the resulting 0-1 knapsack problem attains a surplus $V \leq V^*$. To highlight the interesting case that remains the focus of this paper, we can add the 0-1 restriction as a formal constraint to the problem:

\begin{equation}\label{eq4}
\begin{aligned}
& \underset{x_{1},x_{2},...x_n}{\max}
&& \sum_{i=1}^{n} x_{i}v_{i} \\
& \textit{subject to}: 
&& \sum_{i=1}^{n} x_{i}k_{i}\leq K \\
& \text{and}
&& x_i \in \{0,1\}, \forall i
\end{aligned}
\end{equation}   

\subsection{The knapsack problem as a market design problem} \label{Sec2.2}

In many realistic cases of interest there are multiple agents involved in the process of filling a knapsack; these agents interact strategically to maximize some objective function. Consider, for example, the problem of filling a certain interval of time with TV advertisements. Typically, the TV station is owned by one firm, and the ads themselves are owned by multiple firms. Similarly, in a blockchain there exists a block producer who is in charge of how the knapsack is packed with transactions that are conducted by multiple (usually thousands) of agents. To state this slightly differently, the knapsack is ‘owned’ by one agent, the ‘seller’, who allocates the scarce resource in the knapsack to multiple owners of objects who act as ‘buyers’. This couches the knapsack problem in the familiar setting of a market with buyers and sellers. Moreover, the seller in a market is typically uninformed about vital characteristics of buyers such as the value they place on getting their object packed. In this circumstance, the value $v_i$ is private information of a buyer, and the seller and buyers operate in an environment of incomplete information. The market setting adds multiple layers of complexity to the simple decision-theoretic model described in Section \ref{Sec2.1}.

In any market, the main problem is how goods are allocated to different agents and at what price. In the context of the knapsack problem, the allocation refers to which objects are packed into the knapsack. As such, \textit{allocative efficiency} involves allocating space to agents who value it the most, subject to the capacity constraints of the knapsack. In general, a seller can come up with different pricing rules; for example, the seller may choose to simply charge a fixed posted price from buyers. The existence of a pricing rule has a number of implications for the knapsack problem. First, there is now a bifurcation between the \textit{value} placed by the buyer for obtaining a place in the knapsack, which $\forall i$ is $v_i$ as before, and the \textit{price} paid by the buyer for this, which we denote $p_i$. So, there is a need to differentiate between the total value received by all buyers who find a place in the knapsack, $\sum_{i=1}^n x_i v_i$, and the total amount they collectively pay to the seller,  $\sum_{i=1}^n x_i p_i$, where $x_i \in {0,1}, \forall i \in N$. The former is the surplus generated by filling the knapsack (which accrues to buyers) and the latter is the revenue received by the seller from doing so. It is worth reiterating that maximizing the surplus (and not revenue maximization) is tantamount to achieving allocative efficiency. Moreover, the pricing scheme can (depending on the nature of the scheme) alter the behavior of buyers as they strategically vie to find a place in the knapsack. The elements of the knapsack problem with incomplete information are summarized below. 

\vspace{0.25cm}

\noindent \textit{\textbf{Model II: The general knapsack problem with multiple agents and incomplete information}} 
\begin{enumerate}[nosep, leftmargin=*]
\item A set $N=\{1,2,3,...,n\}$ of \textit{agents}, each of whom owns a single \textit{object}. The notation $i$ is used to denote an object and its owner; 	the context will make it apparent what $i$ refers to. These agents are \textit{buyers} of space in the knapsack.
\item An $(n+1)$th agent who owns (or has rights to fill) the knapsack of capacity, $K>0$. This agent is the \textit{seller} of space in the knapsack.
\item Each object $i\in N$ has a \textit{size} of $k_{i}>0$ associated with it.
\item Each object $i\in N$ has a \textit{value}, $v_i$, associated with it, which accrues to the owner (buyer) when they are successful in purchasing space in the knapsack (and $v_{i}=0$ otherwise). 
\item The seller receives a \textit{payment} (pr price) $p_i$ from buyer $i$ when space is allocated in the knapsack for object $i$
\item The seller and buyers can observe the size $k_i$ of all objects, but the value $v_i$ is private information for $i$. This makes the knapsack problem one of incomplete information.
\item A decision process, $x_i$, that determines which objects are are placed in the knapsack. The interesting case is the binary decision process where $x_{i} \in \{0,1\}, \forall i \in N$; with $x_{i}=1$ when object $i$ is packed in the knapsack and $x_{i}=0$ when it is not. 
\item There are now multiple objectives the seller can have:

\begin{enumerate}[nosep]
\item \textit{Revenue maximization}
\begin{equation}\label{eq5}
\begin{aligned}
& \underset{x_{1},x_{2},...x_n}{\max}
&& \sum_{i=1}^{n} x_{i}p_{i} \\
& \text{subject to} 
&& \sum_{i=1}^{n} x_{i}k_{i}\leq K \\
& \text{and}
&& x_i \in \{0,1\}, \forall i
\end{aligned}
\end{equation}   

\item \textit{Surplus maximization (allocative efficiency):}
\begin{equation}\label{eq6}
\begin{aligned}
& \underset{x_{1},x_{2},...x_n}{\max}
&& \sum_{i=1}^{n} x_{i}v_{i} \\
& \text{subject to} 
&& \sum_{i=1}^{n} x_{i}k_{i}\leq K \\
& \text{and}
&& x_i \in \{0,1\}, \forall i
\end{aligned}
\end{equation}   
\end{enumerate}

\end{enumerate}

It is worth noting that the above model limits the number of objects each agent owns to one. However, a more general setup could consider the possibility of agents with more than one object. Given that information is asymmetric, owning more than one object could potentially change the outcome. For instance, in auctions where the payment is determined by other bids, bidders could have incentives to lower one of their bids, which would reduce their expected payment for the other object.

\subsection{Mechanism design and knapsack auctions} \label{Sec2.3}

The reinterpretation of the knapsack problem as a market situation suggests that the design of block allocation is in some sense similar to the design of a market. Economists are familiar with this; the design of markets and other institutions falls within the broad rubric of mechanism design, where the focus is on specifying an allocation rule to determine $x_i$ for each  $i$ along with a pricing rule that determines how much buyer $i$ has to pay for a given allocation. In general, much of the focus of economic analysis is to design mechanisms that achieve allocative efficiency and elicit truthful information about the private information possessed by each agent. In contrast, computer scientists who investigate this problem in the field of \textit{algorithmic game theory} have dealt with the issue of how to design mechanisms for situations where allocative efficiency is difficult to realize due to the complexity of the problem.\footnote{\cite{roughgarden2016twenty} presents a succinct introduction to algorithmic game theory; for a more detailed exposition, see \cite{nisan2007introduction}.} The knapsack problem is one such situation, where finding a solution to the surplus maximization program is NP-hard even when there is complete information about $v_i$; incomplete information adds a further layer of complexity to the design problem. In such circumstances, algorithmic game theorists have proceeded by sacrificing the goal of surplus maximization to achieve computational tractability. In the triad of desirable goals for the mechanism designer – (i) obtaining truthful revelation of private information, (ii) achieving allocative efficiency and (iii) maintaining computational tractability for practical reasons – economists have focused on (i) and (ii) and by and large ignored (iii), while computer scientists have recognized the importance of forgoing (ii) to achieve (iii) and have investigated the consequence of doing so.  

A popular mechanism in markets, especially when a seller is faced with incomplete information and the lack of enough data to ascertain reasonable prices to charge buyers, is an auction. In a \textit{knapsack auction}, the decision of how to allocate space in the knapsack and the prices to charge for allocated space is determined through an auction where buyers place bids for getting their object packed. Knapsack auctions are multi-unit auctions in the sense that there is more than one item being auctioned, which implies that there can be multiple winners. When allocating transactions to blocks, multiple agents succeed in getting their transactions packed. Different blockchains have adopted different types of auctions. Bitcoin, for example, involves an open auction where each agent can observe the bids (fees offered) by others; Ethereum, on the other hand, has increasingly adopted procedures where sealed-bid auctions are conducted, for reasons related to MEV that we outline in the next section. As such, we focus our attention on sealed-bid auctions in what follows given its importance for MEV. It is important to note that despite the rules of open and sealed-bid auctions being different, they can be strategically equivalent (or outcome equivalent). For instance, it is very well known in the auction theory literature that the sealed-bid first-price auction is strategically equivalent to the open descending auction and the sealed-bid second-price auction is strategically equivalent to the open ascending auction \citep{Krishna2009}.

While the design space for constructing an auction is virtually endless, there are three types of sealed-bid multi-unit auctions that are used in practice and/or have been extensively researched:

\noindent (a) \textit{Discriminatory price (DP) auctions}: Allocation is prioritized in decreasing order of bids, and each agent who wins in the auction has to a pay a price equal to the amount they bid.

\noindent (b) \textit{Generalized second-price (GSP) auctions}: Allocation is prioritized in decreasing order of bids, and each agent who wins in the auction pays a price equal to the amount of the next-highest bid.

\noindent (c) \textit{Uniform-price (UP) auctions}: Allocation is prioritized in decreasing order of bids, and each agent who wins in the auction pays a (uniform) price equal to the highest losing bid. 

The DP and UP auctions are used by numerous governments to auction treasury bills, emissions permits, spectrum licenses and electricity while the GSP auction has gained popularity in selling keywords on internet search engines like Google.\footnote{See \cite{Edelman2007} for an analysis of the GSP auction and \cite{Khezr2020b} for a review of multi-unit auctions.} We examine the usefulness of knapsack auctions for blockchain design in Section \ref{Sec4}.

\section{Maximal extractable value (MEV): the case of Ethereum} \label{Sec3}

\subsection{Value extraction} \label{Sec3.1}

Suppose there exists a mempool where transactions are public. Some of these transactions may be for purchasing goods and services; for example, bitcoin (BTC) or ether (ETH) could be used to purchase a cup of coffee which has some value (net of costs), $v_i$, for the user $i$ placing the transaction.\footnote{In economics, and specifically in microeconomics, this value is often referred to as the \textit{maximum willingness to pay} of a consumer for the purchased good. For a trade to happen the maximum willingness to pay (value) has to be greater or equal to the price paid. Otherwise the consumer has no motivation to purchase the good.} Since this transaction is public, others can see it; however, there is little, if anything, another agent can do to extract this value away from $i$. Frontrunning the transaction, for instance, yields nothing unless the frontrunner has some need to purchase the exact same cup of coffee from the same vendor before $i$ does, a possibility that has no real practical relevance. In this case, the transaction does not present any possibility of \textit{extractable} value for others. 

Now consider an alternative situation where user $i$ sends funds mistakenly to an incorrect smart contract that permits its withdrawal by others who spot the funds in the contract. Consequently, $i$ may be tempted to enter another transaction to withdraw the funds from the smart contract to their own wallet before someone else does. However, this transaction to transfer the funds out of the smart contract is publicly visible to anyone watching the mempool, and so ends up pointing attention towards the funds; another agent, say $j$, could recognize this opportunity and place a transaction to transfer the funds out of the smart contract before user $i$’s transaction goes through. Given a system that prioritizes transactions based on fees for block production, such a (rogue) user $j$ would be able to access the funds before user $i$ does by submitting higher fees to the block producer. In this instance, $i$'s value (equal to the amount of funds in the smart contract) can be extracted by $j$. The situation described here was indeed encountered by \cite{robinson2020}, who liken the Ethereum landscape to a \textit{dark forest}.\footnote{Based on the novel “Dark Forest” by Cixin Liu, where detection of location is tantamount to death by an advanced predator.} On Ethereum, $j$ is typically not a human agent: it is a bot searching the landscape for opportunities to extract value. Entities that scour the mempool landscape looking for opportunities such as this to extract value are labelled \textit{searchers}.\footnote{We note that while in this section our main focus is on the Ethereum ecosystem, the issues raised here are applicable for other blockchains as well.} 

Over time the blockchain community has realized that information about the nature of transactions allows for the possibility of value extraction in various ways when blocks are created. The concept that has been developed to capture this phenomenon is \textit{maximal extractable value} (MEV). The Ethereum website provides a widely accepted working definition:\footnote{See \url{https://ethereum.org/en/developers/docs/mev/}, accessed 6th March 2024.}

\begin{quote}
``\textit{Maximal extractable value (MEV) refers to the maximum value that can be extracted from block production in excess of the standard block reward and gas fees by including, excluding, and changing the order of transactions in a block.}''
\end{quote}

\noindent There are a couple of key ideas in the definition. The first is that MEV occurs during block production. At a very intuitive level, blockchains can be thought of as having three layers: layer 1, which deals with rules for consensus (and hence rules for block production); layer 2, which deals with scaling solutions; and an application layer, which runs various applications such as decentralized finance (DeFi) products, decentralized exchanges, NFT sales and so on. MEV essentially occurs because transactions conducted in the application layer can be manipulated based on rules for block production created in layer 1.\footnote{For this intuition, and much more on how blockchains are constructed and operate, see the Tim Roughgarden lectures: \url{https://www.youtube.com/@timroughgardenlectures1861/videos}, accessed 6th March 2024.} Consequently, there has been considerable effort in altering the block production mechanisms in layer 1 to mitigate the impact of MEV. Secondly, the manner in which MEV is realized is by including, excluding or a changing the order of transactions in a block. The power to extract MEV, therefore, ultimately lies with entities on the blockchain who have the power to manipulate the way transactions are packed into a block. 

There are various types of transactions conducted in the application layer of Ethereum that are susceptible to value extraction during block production. Some examples of such MEV transactions include arbitrage, liquidations, front-running, sandwich attacks, time-bandit attacks and other types of MEV transactions that are lumped together as long-tail MEV. Except the case of arbitrage below, we do not describe all these transactions in detail; the interested reader is referred to the Ethereum website,\footnote{The website maintained by Ethereum on MEV is \url{https://ethereum.org/en/developers/docs/mev/}, accessed 6th March 2024.} \cite{barczentewicz2023mev} or \cite{tuomikoski2023}, all of which contain accessible introductions to MEV in general, and these transactions in particular.\footnote{Apart from the sources listed here, the literature on MEV, both academic and general interest, has expanded considerably as it becomes more important in blockchain design. The classic work on MEV which introduced this idea in a formal way (albeit as miner extractable value instead of maximal extractable value) remains \cite{daian2019flash}.}

We do, however, take a closer look at arbitrage, mainly to highlight a few interesting aspects of MEV. Arbitrage refers to opportunities for obtaining riskless profits due to pricing anomalies in different markets or sets of assets. As a simple example, suppose there are two tokens, $X$ and $Y$, with the price of token $X$ in terms or $Y$ being $P_1$ in one location and $P_2$ in another. If $P_{1} \neq P_{2}$, arbitrage would involve buying $X$ where it is cheaper and selling it in the other location where its price is higher, which results in arbitrage profits equal to the difference between the two prices. Other types of arbitrage such as 3-point arbitrage are also feasible on decentralized exchanges like Uniswap \citep{mohan2022automated}. Indeed, blockchain finance applications such as decentralized exchanges (DEXs) rely on arbitrage to eliminate pricing anomalies that exist between different markets (two-point arbitrage), or internally within the platform (three-point arbitrage). 

There are numerous scenarios related to arbitrage that lead to MEV. One situation, for example, arises when a user spots an arbitrage opportunity and puts in the buy and sell order required to take advantage of the opportunity. A searcher (bot) observing the mempool can copy the same transactions and, with a higher fee, receive priority in the block in terms of the order in which transactions are executed, thereby extracting the value away from the user.  Alternatively, as a second example, the user may \textit{create} an arbitrage opportunity through their transactions (say trading on a decentralized exchange), which can be exploited by a searcher by inserting arbitrage transactions in the block after the user’s actions have been executed. One can observe an important difference between these two examples. In the first example, the user is adversely affected by the searcher’s actions that have essentially taken away the user’s arbitrage profits. This clearly harms the user. Importantly, it harms the blockchain ecosystem in general because arbitrage is necessary to align the prices in many DEXs that employ automated market makers (AMMs); if an arbitrage opportunity spotted by one user is exploited by others, then there is little incentive to put in effort to locate arbitrage opportunities in the first place, which creates inefficiencies in the operation of the AMM.  In the second example, the user’s action creates a price anomaly, which the searcher spots and eliminates by undertaking the arbitrage as soon as the user’s transaction is executed. This does not harm the user, and is in fact good for the ecosystem because the searcher’s actions have aligned prices 
through arbitrage.  Depending on the context, therefore, searcher activity may free-ride of the efforts of others and cause them harm, or in other instances improve the efficiency of applications. This suggests that MEV is a subtle topic that requires careful management and informed policy oversight \citep{barczentewicz2023mev, poux2022maximal}. Figure \ref{fig2} below, which modifies Figure \ref{fig1}, summarizes the impact of MEV.

\begin{figure}[h!]
\centering
  \includegraphics[scale=0.78]{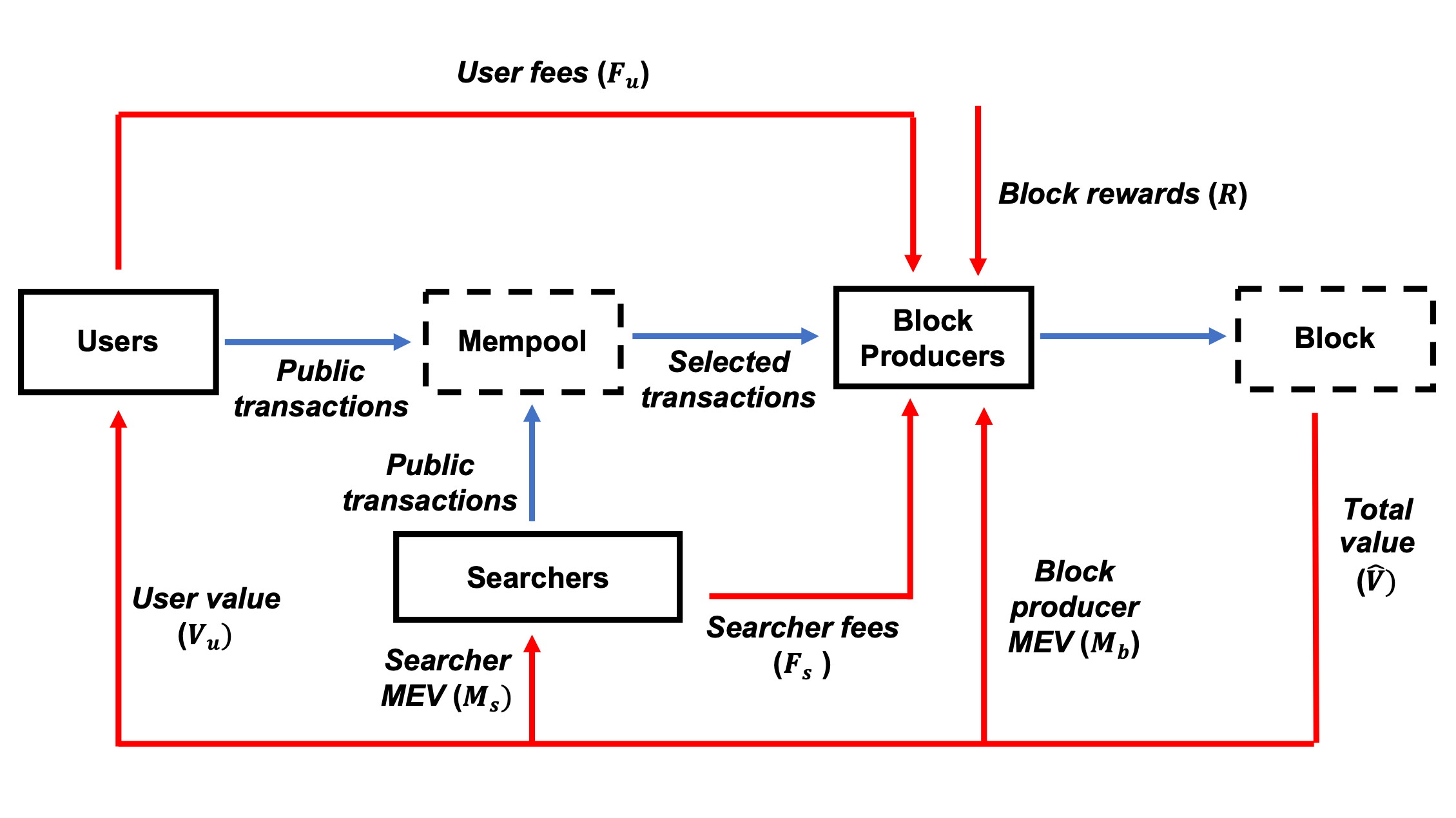}  
\caption{Flows of funds and transactions with MEV}
\label{fig2}
 \end{figure}

As can be seen in Figure \ref{fig2}, relative to Figure \ref{fig1}, the change relates to the presence of searchers who extract information from the mempool and use it either to replace users' transactions with their own, or to create new opportunities for profits. Consequently, some of the value created by the block now accrues to searchers, and to block producers (who can take also advantage of MEV because they have the power to order transactions in a block and essentially become searchers themselves). Searchers receive value not only by \textit{diverting} some value away from users (for example, by front-running user arbitrage transactions with their own) but also by \textit{creating} new value that did not exist before (for example, by back-running user transactions to take advantage of new arbitrage opportunities). It makes sense, therefore, to separate MEV transactions into \textit{value diverting MEV transactions} and \textit{value creating MEV transactions}. Essentially, value diverting MEV makes some user worse off, while value creating MEV does not; indeed, the latter increases the efficiency of the system through the value it adds.\footnote{There is a parallel between this and a customs union, which causes trade creation and trade diversion when it is formed between countries in a regional neighborhood. Trade creation occurs when new trade between member countries takes place due to the elimination of tariffs in the union, which increases efficiency as high-cost producers are replaced by more efficient ones within the union. On the other hand, trade diversion is inefficient and occurs as imports from low-cost producers outside the union are replaced by less efficient production within the union, which becomes feasible due to newly instituted tariffs by the union with the rest of the world.} While our focus has been on arbitrage, one could categorize other MEV transactions into one of these two categories. For example, liquidations undertaken by searchers are value creating, whereas a sandwich attack is value diverting, and so on. Ultimately, whether MEV is, on the whole, desirable or not for a blockchain community depends on which of these two effects is dominant. 

Finally, we can use Figure \ref{fig2} to track the payoffs accruing to the 3 types of agents in our model: users ($\pi_u$), searchers ($\pi_s$) and block producers ($\pi_b$):

\begin{equation}\label{eq7}
\begin{aligned}
& \pi_{u} = V_{u} - F_{u}
&&\\
& \pi_{s} = M_{s} - F_{s} 
&&\\
& \pi_{b} = M_{b} + F_{s} + F_{u} + R 
&&
\end{aligned}
\end{equation}   

\noindent The total surplus from the creation of a block, which is distributed according to Equation \ref{eq7},is:

\begin{equation}\label{eq8}
\begin{aligned}
&\Pi = \pi_{u} + \pi_{s} + \pi_{b} = V_{u} + M_{s} + M_{b} + R = \hat{V} + R\\
\end{aligned}
\end{equation}

\subsection{Privacy and MEV} \label{Sec3.2}

Figure \ref{fig2} represents the state of affairs on Ethereum before 2020, with Ethereum operating on a proof-of-work consensus. Consequently, initial explorations into the MEV problem focused on \textit{miner} extractable value; this terminology was later adapted to replace `miner' with the more general `maximal' to bring proof-of-stake systems (that have validators instead of miners) within the purview of the analysis. At that time, much of the transactions were done through the mempool, with bots acting as searchers and exploiting opportunities for MEV when they were spotted (as shown in Figure \ref{fig2}). 

A key aspect of Figure \ref{fig2} is that searchers, having identified an opportunity for profit, submit their own transactions to the public mempool, with a gas fee that accrues to the block producer if the transaction is packed in a block. These transactions, being public, can be spotted by another searcher, resulting in competition for the profit opportunity as searchers engage in an open-bid first-price auction on-chain via the mempool; these auctions that allocate precedence in the block building ordering are labeled \textit{priority gas auctions} (PGAs; see \citealp{daian2019flash}). This process has a few important negative implications. First, the profits identified by the searchers are competed away in the open-bid PGAs, resulting in the block producers (miners) extracting the entire MEV as rent. Second, since the (failed) transactions by losing searchers were also recorded on-chain, they still incurred a gas cost; this was, therefore, an all-pay auction.\footnote{Indeed, auctions with this structure are typically games of escalation. The classic experiment that displays this is the Shubik dollar auction \cite{shubik1971dollar}.} Third, the PGA activity by bots acting as searchers added to network congestion, which resulted in negative externalities to other users through higher gas costs.\footnote{See the Flashbots documentation at \url{https://docs.flashbots.net/flashbots-auction/overview}, accessed 6th March, 2024; see also Chapter 6 of \cite{tuomikoski2023}.} 

The problems associated with PGAs led to the next development in the ecosystem: the creation of \textit{relays} such as Flashbots that act as intermediaries between searchers and block producers.\footnote{See \url{https://www.rated.network/relays?network=mainnet&timeWindow=30d} for data on relays; accessed 6th March, 2024.} In this scenario, searchers submit transactions to relays off-chain, which maintains the privacy of these transactions by circumventing the mempool. Importantly, the open-bid first price auction conducted in the mempool is replaced by a sealed-bid DP auction in the relay. This prevents a PGA-type escalation of bids to establish priority in the block creation process, thereby ensuring that the issues outlined above are circumvented. The outcome is a move of the ecosystem to Figure \ref{fig3} below. 

\begin{figure}[h!]
\centering
  \includegraphics[scale=0.88]{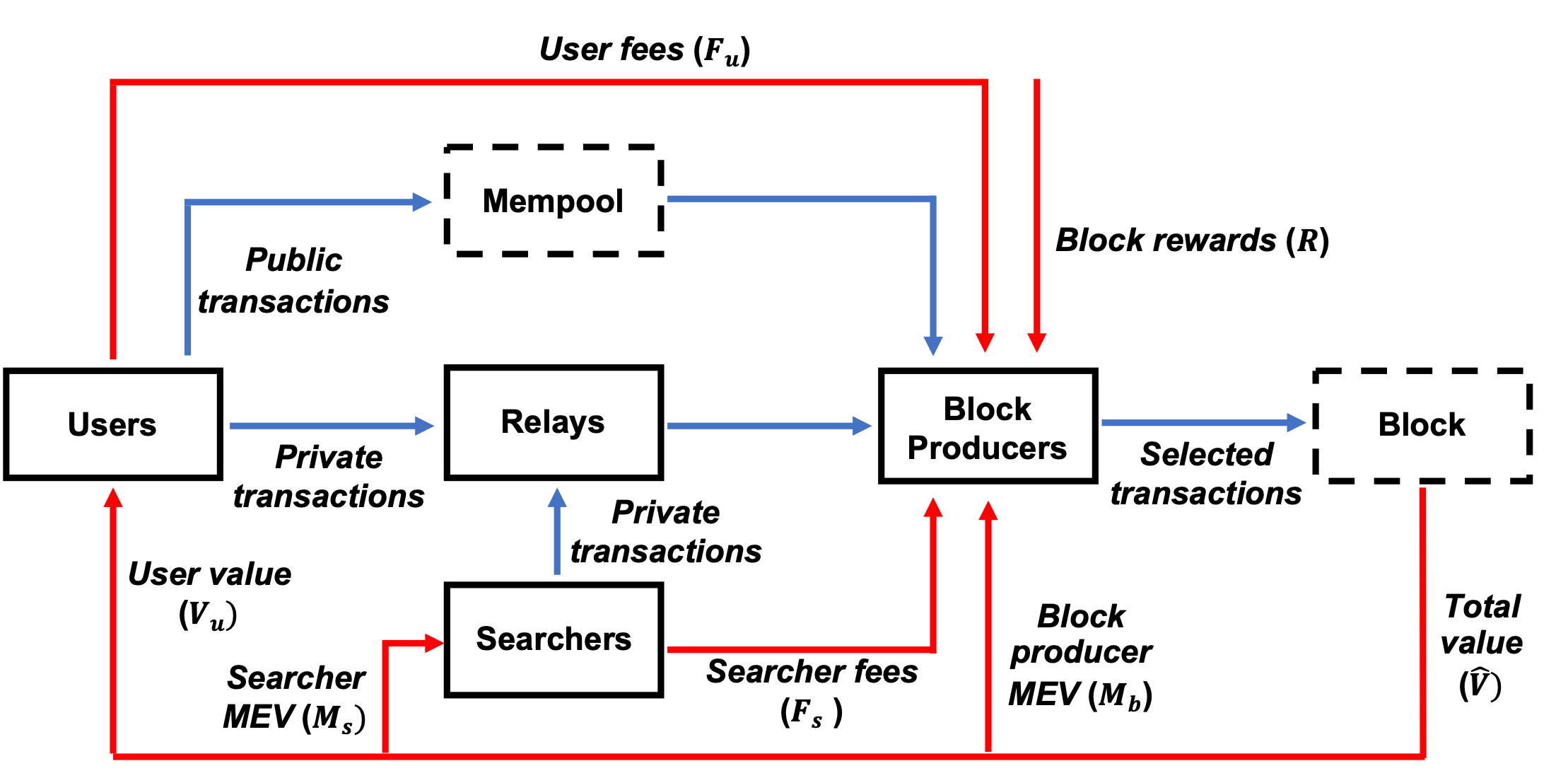}  
\caption{The introduction of relays}
\label{fig3}
 \end{figure}

As can be seen in Figure \ref{fig3}, the main addition is the relay that takes private transactions from searchers and conveys this to block builders, with fees equal to the searchers' bids in the sealed-bid auctions. It is worth noting that relays also offer privacy to users, who can choose to send their transactions to the relay and participate in the sealed-bid auction, rather than send their transactions to the public mempool; this affords users protection against their transactions being front-run or sandwiched. The block producer then converts the inputs, which include now both mempool transactions as well as the transactions transmitted by the relays, to the block output by selecting transactions to fit the block knapsack.

The economics of the relay system is interesting in its own right. In general, the relay is a platform that brings together two sides of the market: the consumers of block space (users and searchers), and block producers.\footnote{See \cite{armstrong2006competition} and \cite{rochet2003platform} for the basic economics of two-sided markets.} As such, one would expect a two-sided network externality where block producers benefit from a greater number of consumers, while at the same time users and searchers benefit from a greater number of block producers in the system to process transactions. In the standard two-sided model, a platform charges fees from one or both sides (depending on the situation) and potentially subsidises one side for participating in the platform. In reality, the economic incentives for relays to operate are not well established at this stage. While making relays a profitable venture through market based pricing is one way to incentivize the entry and development of relays, there are arguments that have been put forward in the community that relays are a public good and need to be funded by the community as such.\footnote{See, for example, this debate between Matt Cutler and Hasu in the Bell Curve podcast: \url{https://www.youtube.com/watch?v=17n6oHoo7pw}, accessed 6th March 2024.} Indeed, many relays that exist at this point in time are funded by venture capital or blockchain foundations, rather through self-sustaining market pricing. The economics of relays is a fruitful area of research.

Assuming zero payments to relays, we can see that Equations \ref{eq7} and \ref{eq8} once again describe the payoffs and surplus in the system (though their magnitudes now change due to the presence of relays and auctions becoming sealed-bid). It is worth emphasizing that if relays are paid in the block creation process, possibly through payments by block consumers or producers, or both, this will redistribute some value towards relays away from other entities.\footnote{We have also assumed that the costs of block building are zero. If block producers and searchers, for example, incur costs for conducting their activities, that is readily factored in to calculate the surplus net of these costs.}

\subsection{EIP - 1559} \label{Sec3.3}

The next significant change in the Ethereum ecosystem occured in August 2021 with the implementation of the Ethereum Improvement Proposal (EIP) 1559.\footnote{See \url{https://eips.ethereum.org/EIPS/eip-1559}, accessed 6th March 2024.} EIP-1559 brought two important changes: (a) alterations to the fee structure; and (b) variable block sizes. We summarize the implications of both of these below; \cite{roughgarden2020transaction} presents an excellent in-depth analysis of EIP-1559 and the mechanism design behind it. 

\begin{figure}[h!]
\centering
  \includegraphics[scale=0.88]{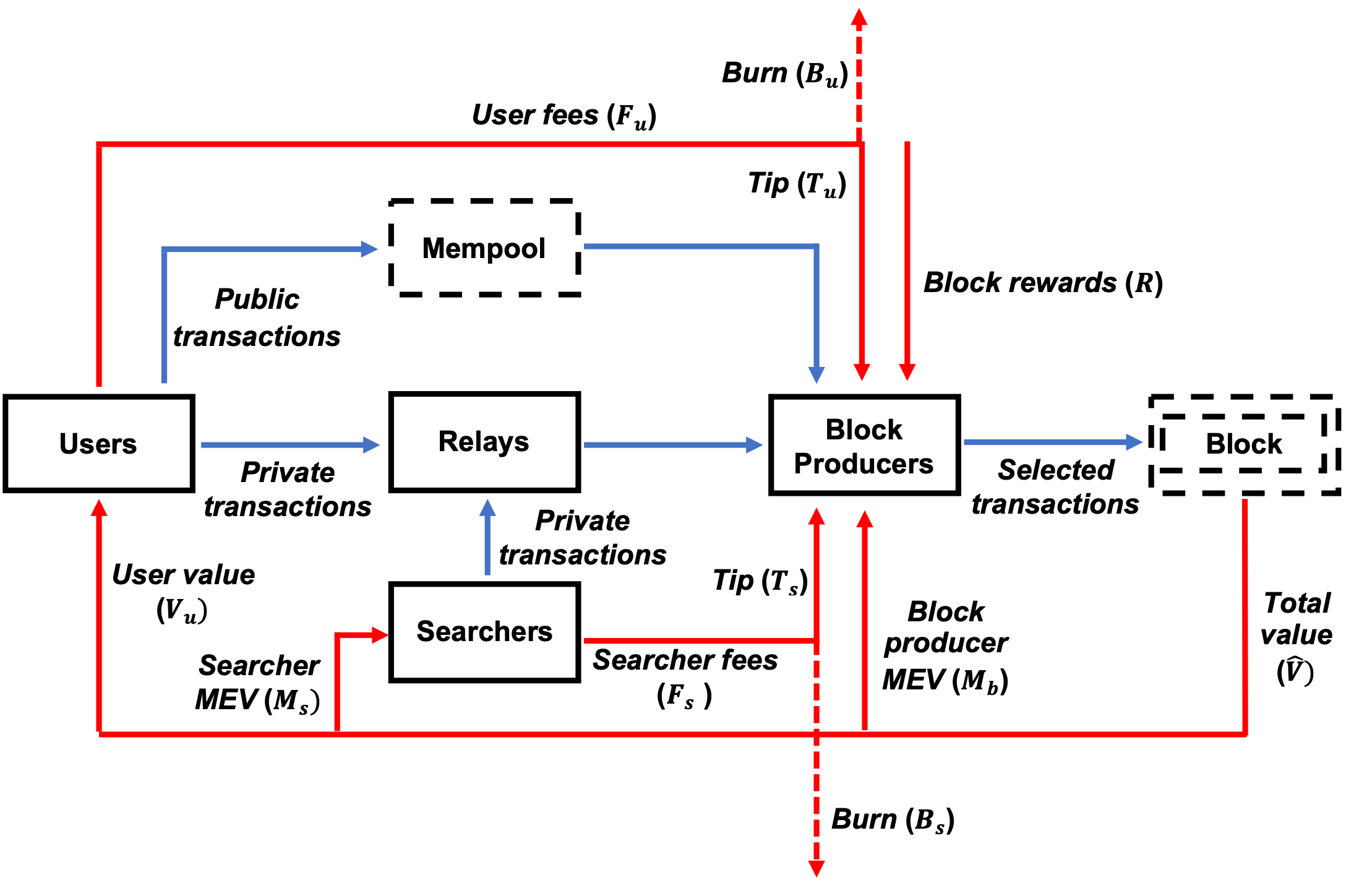}  
\caption{Post EIP-1559}
\label{fig4}
 \end{figure}

One of the characteristics of the Ethereum fee structure prior to EIP-1559 was that it relied on (first-price) auctions, either open (prior to the introduction of relays) or sealed (after the introduction of relays). As optimal bidding in a first-price auction is not trivial, one way to make the fee structure simpler is to have a posted price (per unit of gas) that all transactions must pay to be included in the block, which is referred to as the \textit{base fee}. The base fee acts as a \textit{reserve price} for block inclusion. A user must pay at a minimum, therefore, the base fee per unit of gas multiplied by the gas usage of the transaction to see it included in a block. Once paid, under the EIP-1559 scheme, the base fee is burned in its entirety rather than being handed over to block producers (as one would expect), as this prevents collusion between block producers and users in circumventing the fee.\footnote{See \cite{roughgarden2020transaction} for a game theoretic analysis of this issue.} In Figure \ref{fig4}, this is shown through the dashed lines showing that the aggregate burned funds, $B_u$ and $B_s$, leave the flow of value in the block production process. 

Apart from the compulsory base fee that is burned, users can offer a \textit{priority fee} as part of the gas fees to the block producer; the priority fees can be small or large, depending on the circumstance of how rapidly and in what position the user would like the transaction appear in a block. Moreover, nothing prevents a user from directly transferring funds to a block producer; consequently, we we will assume that the total \textit{tips}  received by the block producer includes both priority fees and direct transfers. Assuming that the aggregate tips paid by users and searchers are $T_u$ and $T_s$ (see Figure \ref{fig4}), respectively, the payoffs to users, searchers and the block producers are now:

\begin{equation}\label{eq9}
\begin{aligned}
& \pi_{u} = V_{u} - F_{u} = V_{u}-B_{u}-T_{u}
&&\\
& \pi_{s} = M_{s} - F_{s} = M_{s}-B_{s}-T_{s} 
&&\\
& \pi_{b} = M_{b} + T_{s} + T_{u} + R 
&&
\end{aligned}
\end{equation}   

\noindent Letting $B_{u} + B_{s} = B$, the total surplus from block creation is now:

\begin{equation}\label{eq10}
\begin{aligned}
&\Pi = \pi_{u} + \pi_{s} + \pi_{b} = V_{u} + M_{s} + M_{b} + R - B = \hat{V} + R - B\\
\end{aligned}
\end{equation}

\noindent We can see from Equation \ref{eq10} that, compared to the situation in Equation \ref{eq8}, the amount $B$ has been lost from the flow of value on the system: it is an amount spent by the consumers of block space that does not flow to any entity in the ecosystem. 

Equation \ref{eq10} points towards some interesting economic features that results from the burning scheme. Firstly, for every block created, $R-B$ measures the net growth of ETH in circulation. Given that $R$ is a fixed amount, but $B$ is an increasing function of the volume of transactions, $N$ (measured in gas), the net growth can be written as $R-B(N)$. It follows that there exists a threshold $\bar{N}$ such that $B(N)>R$ for any $N>\bar{N}$. So, for transaction levels greater than $\bar{N}$ the Ethereum ecosystem becomes \textit{contractionary}, with more ETH leaving circulation than what is injected in the creation of each block. Overall, the burning of the base fee has been likened to a share buyback that increases the value of every unit of ETH that remains \citep{roughgarden2020transaction} and, consequently, a contractionary ETH has been welcomed by many in the Ethereum community. However, this relates to the price of ETH in terms of another monetary unit, say USD. Without reference to such an exchange rate, Equation \ref{eq10} makes it clear that, all else being the same, $R - B < 0$ reduces the total surplus from block creation, as $\Pi$ falls below $\hat{V}$. In other words, any benefits of a contractionary ETH in terms of its price against the USD comes at the expense of a loss in the total surplus from the blocks that are created. Secondly, the fact that the amount burned $B$ is an increasing function of $N$ implies that Ethereum circulation contracts precisely during periods when block space demand by users is high.

In a traditional economy, one can expect the contraction of money supply to have an impact on the general price level of goods and services produced by the economy (measured through something like the Consumer Price Index or GDP deflator). The contraction of money supply, and the concomitant increase in interests rate, generally result in a dampening of inflation (disinflation) or if it is carried too far, negative inflation (deflation). In a blockchain network like Ethereum, it is not clear what the `general price level' is. If one presumes that the main good (or one of the main goods) that ETH is used to pay for is block space, then one could intuit that a falling amount of ETH in circulation to facilitate a rising volume of transactions (measured in gas) would result in the `price' of block space - gas fees in general and base fees in particular - falling as well.\footnote{Such an intuition is supported by the Quantity Theory of Money (in transaction form), which states that $mv = pt$, where $m$ is the amount of money in circulation, $v$ is the velocity of money, $p$ is the price level and $t$ is the volume of goods and services transacted. Assuming $v$ is stable, a rise in $t$ together with a fall in $m$ must be supported by a fall in $p$.} This is perhaps why many in the Ethereum community expected the gas fees to fall with the introduction of EIP-1559. This is, however, an entirely monetarist view that ignores the demand and supply of block space which, in reality, dictates gas fees at any given point in time. One, therefore, needs to be somewhat careful how to interpret the consequences of monetary contraction in this circumstance. 

The main question that arises at this stage is: how exactly are the base fees determined in EIP-1559? The answer is related to what the base fee is attempting to do in this situation. First, as mentioned before, posted prices are easier for users to deal with than figuring out optimum bidding strategies in first-price auctions. So the first use of base fees is to add simplicity to consumer decision-making. Second, and perhaps more importantly, it is an instrument that can be used to influence demand for block space. The Law of Demand dictates an inverse relationship between the price of a good or service and its quantity demanded; consequently, raising (or lowering) the base fee, which is the minimum price for getting block space allocation, reduces (or increases) the quantity of block space that users would demand. When there is congestion in the network and block capacity is exceeded by demand, a rise in base fees eases the demand pressure; on the other hand, when blocks are created with empty space, a fall in the base fees encourages the demand for block space. In EIP-1559, block fees are set algorithmically (based on a given formula) to respond to how full or empty previously constructed blocks are.\footnote{See \cite{roughgarden2020transaction} for details of the formula.} 

This brings us to the second major change in EIP-1559: variable block sizes. Typically, blockchains (including Ethereum prior to EIP-1559) have fixed block sizes; block sizes are capped to ensure that all nodes have the opportunity to process the information related to newly created blocks in a timely manner. With the introduction of EIP-1559, Ethereum now differentiates between the \textit{target block size} (15 million gas) and the \textit{maximum block size} (30 million gas). This is shown in Figure \ref{fig4} with a block output that can have variable size. So, when the previous block has more than the target size of 15 million, the base fees for the current block are adjusted up to bring the size down; similarly, when the previous block has size less than 15 million, base fees are reduced for the current block to encourage more transactions. The end result is a clever process of incremental adjustment of the base fees to ensure that block sizes are always moving towards the target size.  

A number of important implications follow from the introduction of EIP-1559. First of all, the variable block size implies that the impact of the knapsack problem is now limited to situations when the block sizes reach the maximum (30 million gas on Ethereum). Thus, during surges of demand for block space the system once again reverts to a knapsack auction, and it is once again important that the system has an auction mechanism in place that performs well in the knapsack context. Second, while the base fee attempts to achieve an equilibrium block space equal to the target, it does not factor in the order in which the block is packed; in other words, it does not factor in MEV. When searchers vie for position in a block to take advantage of MEV opportunities, the priority fees or tips they pay are once again essentially an auction, albeit without knapsack constraints except when the system is operating at maximum block size and base fees have not yet dampened demand. Consequently, while EIP-1559 makes many improvements to the Ethereum ecosystem, it does not negate the necessity of constructing optimal MEV auctions, with and without a knapsack capacity constraint. 

\subsection{The merge} \label{Sec3.4}

The next stage in the evolution of the Ethereum network occurred with a move from PoW to PoS as the consensus mechanism, an event referred to as \textit{the merge}.\footnote{So named because of a merge between the Ethereum mainnet and the Beacon Chain proof-of-stake mechanism.} The switch to a PoS system is conditioned, among other things, by a desire to reduce the energy consumption in the block creation process that is inherent in a PoW consensus mechanism. In a PoS system, users stake tokens (in the case of Ethereum, 32 ETH) to become validators who are selected (randomly) to propose new blocks on the blockchain; there is no longer a need to solve complex mathematical puzzles to create a block. With the merge, a number of significant changes took place to the block creation process, some of it as a natural consequence of PoS being introduced, but others as a means to tackle the issue of MEV. Our own focus here on the latter, of course, so we skip many of the irrelevant institutional details related to the merge. 

\begin{figure}[h!]
\centering
  \includegraphics[scale=0.79]{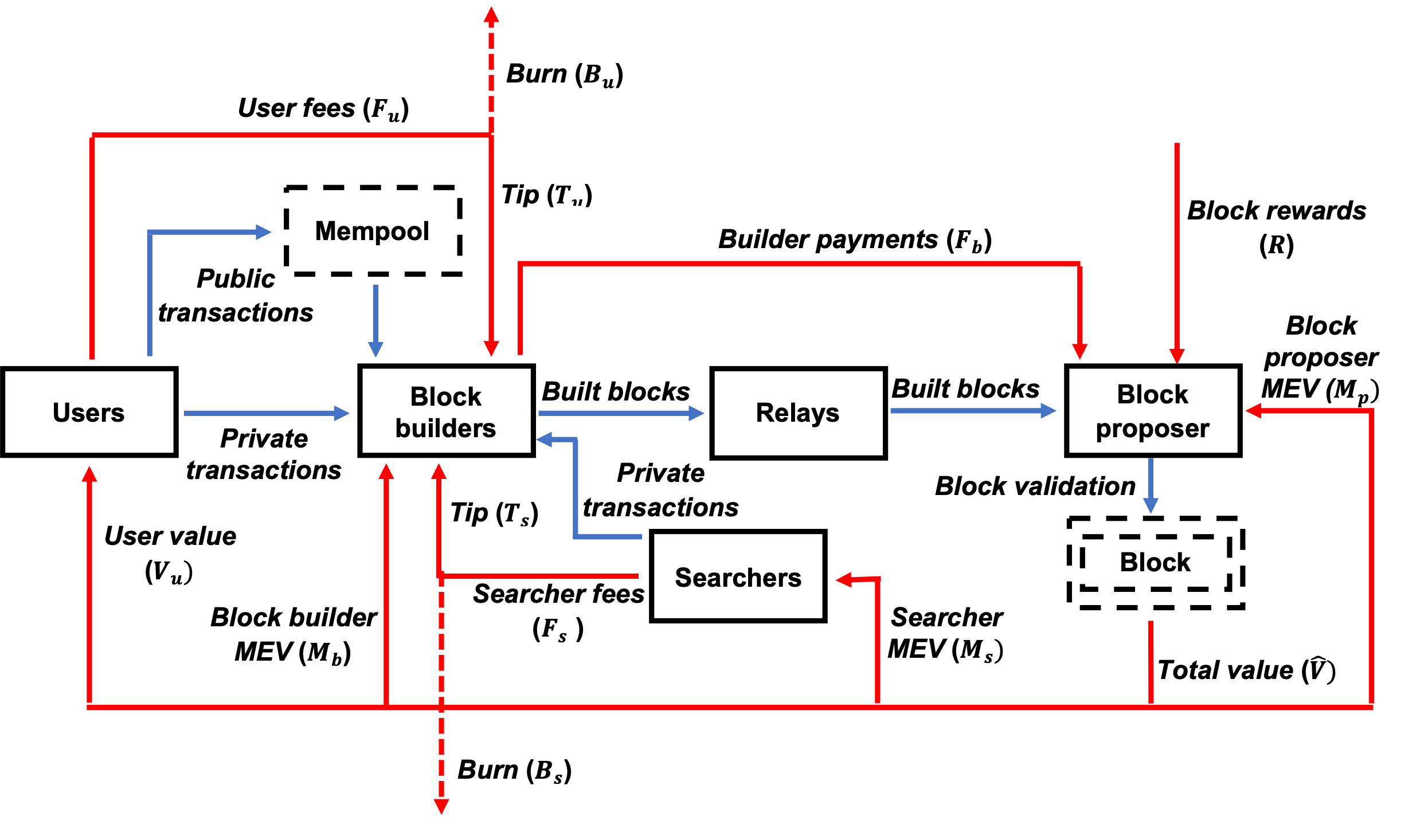}  
\caption{The Merge \& Proposer-builder separation}
\label{fig5}
 \end{figure}

In a PoS mechanism, the role of a block producer, which belonged to the miner in the PoW pre-merge mechanism, is taken over by a validator. Block production involves two activities: first, the creation of a block filled with transaction and second, validating its contents and proposing the block to the network. While the post-merge PoS mechanism technically allows a validator to perform both these activities, the reality of MEV and a desire to limit censorship has resulted in the adoption of a scheme labelled \textit{the proposer-builder separation} (PBS), where these activities are divided into two separate roles performed by a \textit{block builder} and a \textit{block proposer}. This is shown in Figure \ref{fig5}, where the blue lines again track the `real' part of the blockchain system with transactions being converted to blocks. 

Given that the PBS system is in place, users transmit their transactions publicly through the mempool or privately to block builders directly. Similarly, searchers, who extract MEV possibilities from the mempool, submit their transactions privately to block builders; indeed, searchers can submit a \textit{bundle}: a group of transactions that are executed in the same block (in the order prescribed by the searcher) in order to ensure the realization of MEV. Block builders collect public and private transactions and create candidate blocks. These blocks are sent to relays, who act as middlemen between block builders and proposers.\footnote{This role of relays is facilitated through an implementation called \textit{MEV Boost}; see \url{https://github.com/flashbots/mev-boost} for details, website accessed 6th March 2024.} Essentially, relays serve as custodians of the information contained in the block by checking the validity of the block, keeping the contents private and passing along only the header information to the proposer. Since a block proposer may be connected to several relays, the proposer selects the most profitable block, signs the header and returns it the corresponding relay. The full block contents are then sent to the block proposer, who then broadcasts it to the network. Unlike the pre-merge PoW block creation process where the rate at which blocks are created is probabilistic and depends on the time it takes for miners to solve the mathematical puzzles, in PoS the block creation timeline is divided into an \textit{epoch} that has 32 \textit{slots} of 12 seconds each (so that each epoch lasts 6.4 minutes). A validator is selected randomly to propose a block in each slot, which is ratified by a randomly selected committee of validators, who \textit{attest} to the validity of the block.\footnote{See \url{https://ethereum.org/en/developers/docs/consensus-mechanisms/pos/}, accessed 6th March 2024.}

Moving to the monetary and value side of the Figure \ref{fig5}, users and searchers pay base and priority fees for their transactions. As before, following EIP-1559, base fees ($B_u$ and $B_s$) are burnt and represent a leakage in the value flow between agents in the system. The priority fees now accrue to block builders, who may also receive directs transfers from users and searchers; together these payments received by the block builders are marked as tips $T_u$ and $T_s$ in the figure. These tips represent the revenue from block building. Since more than one block builder exists, and each may receive a different set of private transactions, block builders compete with each other by offering payments or \textit{bribes} to block proposers if their block is selected, with relays acting as middlemen to ensure that proposers do in fact receive their promised payments. In Figure \ref{fig5}, the payment from a successful block builder to the proposer is $F_b$. These represent costs for the block builder and revenues for the validator. However, these are not the only source of revenues for a validator; the block proposer also receives the block rewards, which in the post-merge lexicon is referred to as \textit{issuance} (and labelled $R$ in Figure \ref{fig5} for consistent notation with earlier figures).\footnote{See \url{https://ethereum.org/en/roadmap/merge/issuance/}, accessed 6th March 2024.} Unlike the pre-merge PoW system, where miners received a fixed block reward,\footnote{The reward was 2 ETH following the Constantinople upgrade in 2019.} the PoS mechanism calculates issuance using a more complicated formula, with validators receiving (smaller than before) rewards for block proposal and attestation.\footnote{According to \url{https://ethereum.org/en/roadmap/merge/issuance/} (accessed 6th May 2024), the amount of new ETH issued for block creation reduced from 13,000 ETH / day during pre-merge PoW to 1,700 ETH / day under PoS. The rationale for lower payments to validators under PoS is that the energy expence borne by miners in PoW is no longer required in a PoS regime.} Based on this discussion and Figure \ref{fig5}, we can track the payoffs received by the four agents in the system: users ($\pi_u$), searchers ($\pi_s$), block builders ($\pi_b$) and block proposers ($\pi_p$). We note that builders and proposers, like searchers, can independently extract MEV during the block building and proposing activities, and have labelled $M_b$ and $M_p$ in Figure \ref{fig5} to account for this possibility. We also note that, as before, the payoffs and economics of relay operation remain unclear in this environment as well; consequently we abstract away from relay remuneration, while noting that if transfers to relays are important, they can be readily tracked in Figure \ref{fig5}.

\begin{equation}\label{eq11}
\begin{aligned}
& \pi_{u} = V_{u} - F_{u} = V_{u}-B_{u}-T_{u}
&&\\
& \pi_{s} = M_{s} - F_{s} = M_{s}-B_{s}-T_{s} 
&&\\
& \pi_{b} = M_{b} + T_{s} + T_{u} - F_{b}
&&\\
& \pi_{p} = M_{p} + F_{b} + R 
&&
\end{aligned}
\end{equation}   

\noindent As before, letting $B_{u} + B_{s} = B$, the total surplus from block creation is:
\begin{equation}\label{eq12}
\begin{aligned}
&\Pi = \pi_{u} + \pi_{s} + \pi_{b} + \pi_{p} = V_{u} + M_{s} + M_{b} + M_{p} + R - B = \hat{V} + R - B\\
\end{aligned}
\end{equation}

\section{Blockchain knapsack} \label{Sec4} 

The knapsack problem is the fundamental issue in block-building. In this section, we discuss various applications – real and hypothetical – of the knapsack problem in blockchains. Since this is a nascent field, especially in the context of MEV, our goal is to build intuition for how the knapsack problem needs to be adapted in various contexts, and to provide a framework for research in this area. 

\subsection{Production or throughput efficiency} \label{Sec4.1}

Figure \ref{fig1} depicts the primary function of block production: to take as inputs a set of transactions that are waiting to be validated, and convert them to blocks of validated transactions as outputs. At a very fundamental level, one can enquire whether the conversion of inputs to outputs is taking place efficiently, an idea that economists refer to as \textit{production efficiency}, but in the context of digital information (transactions) could be labelled \textit{throughput efficiency} as well because it indicates how fast transactions can move through the system. It is worth noting production efficiency does not factor in how transactions are valued by users; rather, its focus is on getting data converted from input to output as efficiently as possible. This yields the maximization program described in Equation \ref{eq13} below:

\begin{equation}\label{eq13}
\begin{aligned}
& \underset{x_{1},x_{2},...x_n}{\max}
&& \sum_{i=1}^{n} x_{i}k_{i} \\
& \text{subject to} 
&& \sum_{i=1}^{n} x_{i}k_{i}\leq K \\
& \text{and}
&& x_i \in \{0,1\}, \forall i
\end{aligned}
\end{equation}   

\noindent In the knapsack literature, this is often referred to as the subset sum problem, which is still an NP-hard problem \citep{garey1979computers}. \cite{kellerer2004multidimensional} discuss a number of solution techniques, including polynomial time approximation schemes. 

Since production efficiency only factors in transaction sizes (and not values), there exists no private information in this case, and consequently no requirement for an auction or any other procedure to discover user willingness to pay. As such, it represents perhaps the simplest level of efficiency that a blockchain achieves in its block production procedure. The main problem with this, of course, is that it does not necessarily achieve allocative efficiency in the sense of maximizing the surplus realized from each block. Nevertheless, it does allow the mempool to be cleared as rapidly as possible. 

\subsection{Allocative efficiency and revenue maximization in block production} \label{Sec4.2}

When values are taken into consideration, we are in a situation characterized by incomplete information, the elements of which are described in Model II of Section \ref{Sec1}. The mechanism designed to sell block space to buyers in blockchians must now come up with a pricing and allocation rule that deals with private valuation of buyers for the block space. A knapsack auction offers a simple method to sell block space in this circumstance. With a knapsack auction, each bidder places a bid (say in ETH), which we denote $B_i$. Based on the profile of bids $(B_1,B_2,...,B_n )$ and sizes $(k_1,k_2,...,k_n)$ the knapsack auction mechanism provides an \textit{allocation rule} for determining which objects are included in the knapsack and a \textit{price} that each winner has to pay the auctioneer for getting their object into the knapsack.  

Suppose bids are truthful ($B_{i} = v_{i}$); then, surplus maximization requires:

\begin{equation}\label{eq14}
\begin{aligned}
& \underset{x_{1},x_{2},...x_n}{\max}
&& \sum_{i=1}^{n} x_{i}B_{i} = \sum_{i=1}^{n} x_{i}v_{i}\\
& \text{subject to} 
&& \sum_{i=1}^{n} x_{i}k_{i}\leq K \\
& \text{and}
&& x_i \in \{0,1\}, \forall i
\end{aligned}
\end{equation}  

One of basic insights of algorithmic game theory is that in a knapsack auction described by Model II, allocative efficiency (surplus maximization) is an NP-hard problem \citep{roughgarden2016twenty}. Consequently, the practicality of computational tractability requires the use of approximations. A computationally simple allocation rule is the Greedy algorithm \citep{roughgarden2016twenty}, which has the following steps in a blockchain:

\vspace{5pt}

\noindent \textit{Step 1}: Order the bid per unit size $(B_{1}/k_{1},B_{2}/k_{2}, \ldots, B_{n}/k_{n})$; without loss of generality, suppose that $B_{1}/k_{1} > B_{2}/k_{2} > \ldots > B_{n}/k_{n}$.

\noindent \textit{Step 2}: Fill the block with transactions in descending order of the bids per unit size till there is no further space to fill the next transaction.

\noindent \textit{Step 3}: Calculate the surplus (sum of bids) arising from this Greedy algorithm and compare it with the highest bidder not included, and pick the option that generates the higher surplus.

\vspace{5pt}

Step 3 is included to eliminate perverse outcomes when there is a very large bidder. For example, consider again the the example provided in footnote 10, with two objects and a knapsack of capacity $K$. The first object has $v_{1}=1$ and $k_{1}=1$; the second has $v_{2}=K - \delta$ and $k_{2}=K$, where $K - \delta > 1$. Assuming truthful bids, we have that $B_{1}/k_{1} = 1$ and $B_{2}/k_{2} =(K-\delta)/K < 1$; with the Greedy algorithm, the first object is packed but not the second one (as packing both exceeds capacity), which is inefficient. With Step 3 in place, not only is this possibility eliminated, but additionally it can be shown that the surplus arising from the Greedy algorithm is at least half the maximum possible surplus, $V^*$ \citep{nisan2007introduction, roughgarden2016twenty}.

There are a number of known results related to the use of the Greedy algorithm for knapsack auctions, which we summarize below.

\vspace{5pt}

\noindent \textbf{Result 1}: While the Greedy algorithm is computationally simple, it is not necessarily allocatively efficient \citep{nisan2007introduction}. 

\vspace{5pt}

\noindent \textbf{Result 2}: Given that the Greedy algorithm is monotonic (that is, given the bids of other agents, if the item of an agent is packed, then by increasing their bid they should be packed again), there exists a unique auction mechanism that results in truthful bidding \citep{myerson1981optimal}. In other words, there exists a unique pricing rule for the Greedy algorithm such that all agents bid their true values.

\vspace{5pt}

\noindent \textbf{Result 3}: The unique pricing rule that results in truthful bidding involves every winning bidder paying some critical amount below which the bidder loses and above which the bidder wins \citep{nisan2007introduction, roughgarden2016twenty}. This unique pricing rule corresponds to the uniform price (UP) auction, where each winning bidder pays the highest losing bid \citep{buterin2018, khezr2024knapsack}. The main problem with the UP auction is that the mechanism is not credible in the sense that the agent responsible for implementing the mechanism may deviate from the prescribe allocation; indeed, \cite{akbarpour2020credible} show that when considering sealed-bid static auctions, the only credible auction is a first price auction. In the context of blockchains, this implies that a UP auction is susceptible to fake bids by the block producer or to collusion between the block producer and some users sending transactions \citep{buterin2018, roughgarden2020transaction}.

\vspace{5pt}

\noindent \textbf{Result 4}: Since the pricing rule that results in truthful bidding (UP) is unique, auctions with other pricing rules such as discriminatory price (DP) and generalized-second price (GSP) auctions do not result in truthful bidding in the knapsack auction. Moreover, as the Greedy allocation only approximates the maximum surplus, the Vickery-Clarke-Groves (VCG) mechanism is no longer necessarily truthful \citep{lehmann2002truth, mu2008truthful, nisan2007computationally}.

As outlined in Section \ref{Sec3}, in order to address the issues related to MEV, Ethereum has constructed a system that has moved transactions away from the public mempool to privately transmitted channels, with a DP auction to allocate space in blocks. This does not yield a truthful equilibrium, so one may well question what the properties of this auction and how it compares to the truthful equilibrium generated by UP. Similarly, while not used in any blockchain, GSP auctions have found popularity in internet sponsored search auctions \citep{Edelman2007}. Can it be useful in knapsack auctions? Due to the complexity of the knapsack auction mechanism and a paucity in the literature in using GSP in contexts other than sponsored search, theoretical investigations have not yet yielded any clear-cut results. However, to gain some insight on this issue, \cite{khezr2024knapsack} use experiments with human subjects as well as simulations with AI agents to compare the three auction types – DP, GSP and UP – across the two criteria of the revenue ($\sum_{i=1}^{n} x_i p_i$) and the surplus ($\sum_{i=1}^{n} x_i v_i$) generated by each. Their findings are summarized below.

\vspace{5pt}

\noindent \textbf{Result 5}: In the laboratory experiments conducted with human subjects and the Greedy algorithm as the allocation rule, \cite{khezr2024knapsack} report that UP achieved the highest level of allocative efficiency in terms of packing the knapsack, followed by GSP and DP in that order. On the other hand, DP achieved the highest revenue among the three, followed by GSP and UP. AI simulations using a Q-learning algorithm provided results that largely aligned with the experimental outcomes, attesting to the robustness of these rankings.

\vspace{5pt}

Result 5 is instructive in light of the fact that DP auctions are by far the most common auction type used in the blockchain space. The theoretical results of \cite{akbarpour2020credible} imply that the DP auction is credible, which adds to its attraction; however, the UP auction is truthful. The experimental results in \cite{khezr2024knapsack} indicate that DP auctions perform well at generating high revenues (for validators and block builders), but their performance in terms of allocative efficiency is relatively poor. This suggests that blockchain protocols need to assess their objectives carefully: is the primary objective to generate high revenues in order to incentivize validators and block builders (and thereby maintain security of the blockchain), or is it to ensure that users are allocated block space as efficiently as possible given a Greedy algorithm? DP auctions perform well for the former and UP for the latter. However, we suspect that many in the blockchain community will aver that both are equally important. A mechanism that ignores allocative efficiency runs the risk of creating block allocation outcomes that are dissatisfying for the blockchain community, whereas one that ignores adequate revenue creation for block production runs the risk of not incentivizing the consensus and security structure of a blockchain. If that is the case, the GSP auction is possibly the most attractive; the experiments and simulations run by \cite{khezr2024knapsack} suggest that a GSP knapsack auction generates revenues that are very close to DP auctions and achieves a level of efficiency very close to UP auctions. In many ways, this answers the question of why GSP is a useful auction despite being not truthful and is relatively more complicated for users to understand. The use of GSP in the knapsack auctions in general and blockchains in particular is a fruitful avenue for future research. 

\subsection{MEV and the knapsack problem} \label{Sec4.3}

One of the main ideas arising from the discussion in Section \ref{Sec4.2} is that the currently popular DP auction appears to perform poorly in terms of efficiency compared to other commonly used multiunit auctions, such as GSP and UP. There is scope for improvement in the current system even in this basic feature of running knapsack auctions to allocate block space. Efficiency is a vital component in the search for a solution to the MEV problem because an inefficient system runs the risk of prioritizing transactions that exploit MEV at the expense of other transactions in the mempool that have higher value to users. Moreover, any attempt to automate the block building process without control and censorship by miners, validators or other network participants would require a relatively efficient mechanism to exist. An inefficient mechanism always runs the risk of rent extraction by some participant at the expense of another. Consequently, we would argue that moving away from the DP auction towards a UP or GSP auction may be ultimately required as a solution to the MEV problem. Indeed, as outlined above, GSP auctions offer a potential compromise between revenue generation and efficiency, but have yet not been implemented by any blockchain in the block building process.

In the knapsack auctions considered thus far, objects are allocated to the knapsack in any order; objects simply need to be packed to generate full value to the user and bidding behavior does not place any strategic importance to the order in which objects are fitted into the knapsack. Yet, we know from Section \ref{Sec3} that MEV is closely linked to prioritizing certain transactions over others for execution on the blockchain. Under the current system in Ethereum, this is organized through sealed-bid bids in a DP auction regime. This raises two questions for finding ways to manage MEV: first, what are the characteristics of a basic (single agent) knapsack problem with complete information where the order in which objects are placed in the knapsack affects value; second, when there is incomplete information, how does one construct a knapsack auction where the position in the knapsack matters?

Consider the first problem, which involves modifying Model I to allow for position in the knapsack to affect the value of a transaction and results in a position-dependent knapsack problem. \cite{gawiejnowicz2023knapsack} examine this problem when $v_i (r)=v_i f(r)$, where $r$ denotes the position of an object in the sequence of objects packed and $f(r)$ is a function that scales the value $v_i$ depending on the position. They propose a pseudo-polynomial exact algorithm and a fully polynomial-time approximation scheme when $f(r)$ is montonically non-decreasing or non-increasing. As the authors note, position dependent knapsack problems are a new class of knapsack problems for which there are many unanswered questions that need to be addressed in future research, including what happens if the monotonicity assumption is dropped or valuation functions other than $v_i (r)=v_i f(r)$ are considered. 

In the context of blockchains, Model II has to be modified so that bids placed by users not only determine which transactions are included in a block, but also the order in which they are included. There are two different strands of the auction literature that deal with position dependency. The first relates to sponsored search auctions \citep{Edelman2007, aggarwal2006knapsack}, where the highest bid is allocated to the first advertisement spot when results to an internet search are displayed, the second spot to the second highest bid and so on, till the space for ad spots reaches capacity. Sponsored search auctions have been examined both from the point of view of GSP \citep{Edelman2007} as well as a generalized first price auction \citep{Han2015}. The second strand deals with limited-edition auctions \citep{khezr2021property}, where objects with a fixed set of editions (such as artwork and fashion luxury items) are numbered and the order of the bid determines the edition number received by winners. \cite{khezr2021property} show that, unlike the case with knapsack auctions, UP auctions are not truthful in limited-edition auctions. It is not clear at this stage which of these strands will prove more useful in ordered auctioning of block space. Consequently, there is considerable research potential in the design of knapsack auctions with ordering.

\section{Conclusion} \label{Sec5}

This paper has attempted to integrate and outline a number of different ideas: the role of the knapsack problem in block-building, and the changes to blockchains in general, and Ethereum in particular, that have been necessitated by the MEV problem. A comparison of Figure \ref{fig1} with Figure \ref{fig5} serves to illustrate how deep these changes have been. Indeed block building is now much more complicated than it was five years ago, with actors such as searchers, relays, block builders and block proposers emerging in the new landscape. In order to highlight various interrelationships and to keep track of changes in the system over time, especially as it relates to MEV, our paper tracked both the `real' side of the block creation process - the conversion of transactions (inputs) to blocks (outputs) - as well as the monetary and value flows occurring between participants. 

Despite the changes that have taken place, it would seem that there is still a long way to go. \cite{heimbach2023ethereum}, for example, report that PBS tends to stimulate censorship rather than alleviate it, that there exists significant centralization with builders and relays, and that relays may often prove unreliable. More layers of complexity, it seems, simply creates a newer set of problems to contend with. Consequently, we have taken a step back to highlight an important aspect of achieving efficiency in the block production process, especially with MEV as an important factor: careful auction design. While the mechanism design literature in this area is still scarce, this paper has attempted to consolidate the main ideas, primarily to serve as a resource for researchers and students interested in understanding and exploring this area further.   

\clearpage
\bibliographystyle{econ}
\bibliography{MEV}
\end{document}